# Feature-aggregated spatiotemporal spine surface estimation for wearable patch ultrasound volumetric imaging

Baichuan Jiang[a], Keshuai Xu[a], Ahbay Moghekar[b], Peter Kazanzides[a], Emad Boctor[a]

[a]Department of Computer Science, Johns Hopkins University, Baltimore, USA; [b]Department of Neurology, Johns Hopkins Medical Institute, Baltimore, USA

## ABSTRACT

Clear identification of bone structures is crucial for ultrasound-guided lumbar interventions, but it can be challenging due to the complex shapes of the self-shadowing vertebra anatomy and the extensive background speckle noise from the surrounding soft tissue structures. Therefore, we propose to use a patch-like wearable ultrasound solution to capture the reflective bone surfaces from multiple imaging angles and create 3D bone representations for interventional guidance. In this work, we will present our method for estimating the vertebra bone surfaces by using a spatiotemporal U-Net architecture learning from the B-Mode image and aggregated feature maps of hand-crafted filters. The methods are evaluated on spine phantom image data collected by our proposed miniaturized wearable "patch" ultrasound device, and the results show that a significant improvement on baseline method can be achieved with promising accuracy. Equipped with this surface estimation framework, our wearable ultrasound system can potentially provide intuitive and accurate interventional guidance for clinicians in augmented reality setting.

**Keywords:** Bone surface estimation, deep learning, volume reconstruction, wearable ultrasound, lumbar puncture

## 1. INTRODUCTION

Ultrasound imaging has been introduced in recent years to guide clinicians to perform spine interventions such as lumbar puncture, but the complex shapes of the self-shadowing lumbar vertebrae make it challenging for clinicians correctly interpret the images, thus reducing the intervention success rate[1]. The situation is aggravated when combined with other risk factors such as obesity[2] and abnormal vertebrae[3], leading to failures and repeated needle insertion attempts which increases the risk of complications including CSF leak and headache[4]. Therefore, in our earlier work[5], we presented a wearable patch-like ultrasound device solution "AutoInFocus" to allow volumetric imaging of patient lumbar vertebrae from multiple imaging angles for improving the vertebrae surface visibility. Ultimately, the imaging device will be incorporated into an augmented reality setting and aid the clinicians performing spine intervention.

The key idea of the "AutoInFocus" setup is to use a miniaturized robotic mechanism moving a phased array transducer in a two-degree-of-freedom (2-DOF) workspace and imaging the lumbar vertebrae from multiple angles, as shown in Fig. 1. Because the bone surface reflection signal is highly dependent on the incoming beam angle, a multi-angle volumetric imaging setup can potentially collect images for each piece of vertebrae surfaces at appropriate angles. In this paper, we will present our method for estimating the vertebrae surface locations in the ultrasound images acquired by the patch device setup.

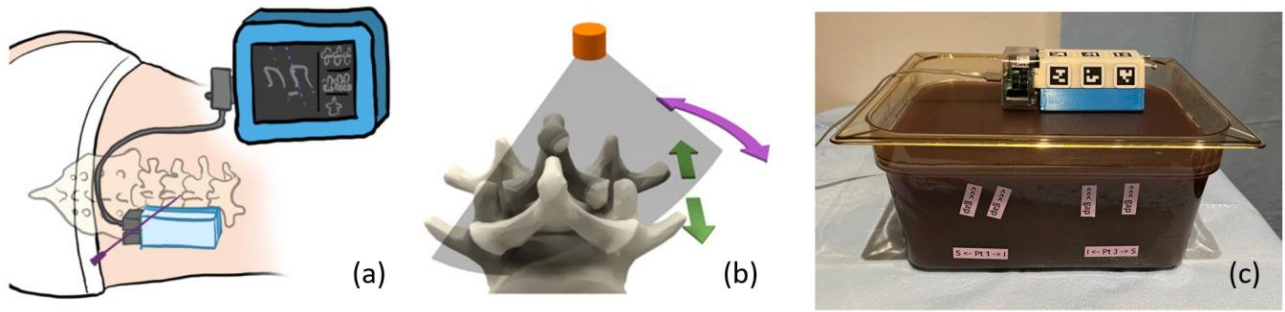

**Figure 1**. Wearable patch ultrasound scanner setup (a) proposed solution of patch device-based lumbar puncture guidance. (b) equivalent 2-DOF workspace with rotation and translation. (c) Data acquisition experiment setup on phantoms.

Given the high noise level and various imaging artefacts, it remains a challenging task to correctly identify vertebrae surfaces. Berton *et al.*[6] leverage random forests-based feature extraction to segment the spinous process in the cross-section view images, but it is limited to only delineating the top part of the spinous process where high reflection is typically observed. Wang et al.[7] instead utilizes a U-Net type neural network combined with filter-based feature maps to estimate bone surfaces, but have not demonstrated its application to the highly complex vertebrae anatomy.

In this work, we propose to utilize a spatiotemporal U-Net architecture with raw image and aggregated hand-crafted feature map as inputs, along with a visibility-based ground truth label generator, to estimate the visible surfaces to the maximum extent based on the image information. With estimated surface locations and device kinematics readings, we show that a 3D volumetric representation of the vertebrae can be successfully reconstructed for interventional guidance.

## 2. METHOD

### 2.1 Aggregated bone surface feature map generation

Before training our backbone neural network, we decided to implement a hand-crafted feature extractor to provide high quality input to the U-Net estimator. This feature extraction workflow is demonstrated in Fig. 2. Given the original B-Mode phased array image input, we first scan converted it back to the polar coordinate image, and two feature extractors are executed and aggregated as final output.

*Phase feature extraction*: The high intensity responses from the tissue-bone boundary is one prominent feature to localize the bone surface. Researchers have proposed using local phase image information to extract the bone boundary based on its symmetry signature[8]. Specifically, a set of 2-D Log-Gabor filters were used and each computes the difference between odd and even filter responses. Intuitively, a narrow bar feature (ridge) in the image could be defined as any point at which all the spectral components have phase congruency at 90°, thus the highest even filter response. An example result is shown in Fig. 2 (green box), where bar features in the image are generally getting high responses.

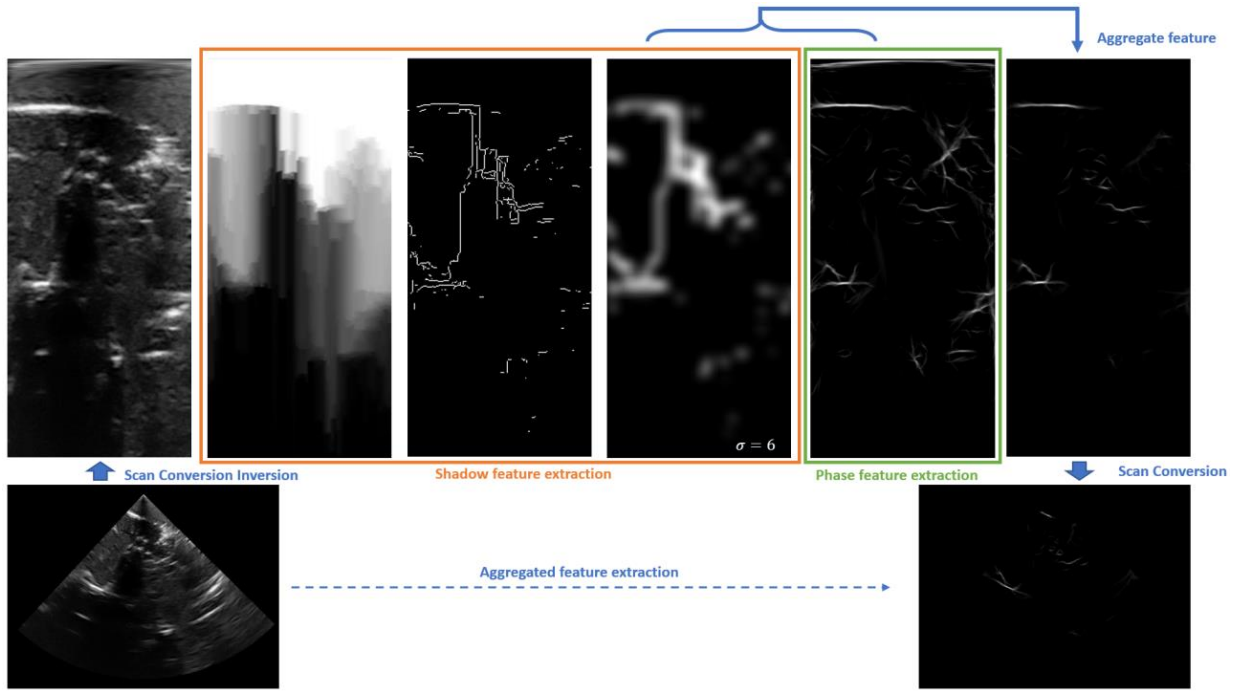

**Figure 2**. Aggregated feature map generation workflow. Original B-mode image from a spine phantom is first scan-converted back into polar coordinate space, and shadow detection and edge detection are executed and aggregated into a single feature map to be fed into a neural network along with the original image as the input.

*Shadow feature extraction*: Shadow is another determining feature for bone structures and it can differentiate bones from other bar feature-inducing reflectors such as muscle, tendon and background speckle noise. Here we implemented a confidence map[9] based shadow feature extractor. The confidence map denotes the information uncertainty in the attenuated (shadow) region, and it can be solved by modeling the problem within a random walk framework: for a pixel in the output feature map, its intensity corresponds to the probability of a random walk starting from this pixel to reach the virtual transducer element, under user defined walking constraints and penalty values. One example can be seen in Fig 2 (orange box): the leftmost figure is the confidence map; and we use Sobel edge detector and a pre-defined threshold value to generate a binary bone boundary map, as in the middle figure; after applying a Gaussian filter with kernel size 6 we get the image on the right.

As it can be observed that the output of the phase symmetry feature extractor is usually noisy due to the misleading bar features present in the image, we propose to aggregate both the phase symmetry feature map and the shadow feature map by element-wise multiplication to achieve mutual agreement between two feature maps. The output feature map is further scan-converted into cartesian space as one input channel to our spatiotemporal U-Net.

## 2.2 Spatiotemporal U-Net with image and aggregated feature channel input

The lumbar vertebrae have a repetitive anatomical appearance pattern longitudinally, and our patch ultrasound device is also scanning in a back-and-forth pattern with a slightly different angle for each sweep as shown in Fig. 1(b). Therefore, to capture this temporal pattern, we propose to incorporate a recurrent neural network (RNN) into a U-Net type segmentation model for spatiotemporal context learning. Inspired by VesNet[10], the model architecture is shown in Fig. 3 below. The two input channels consist of the original B-Mode image channel and the aggregated feature map channel.

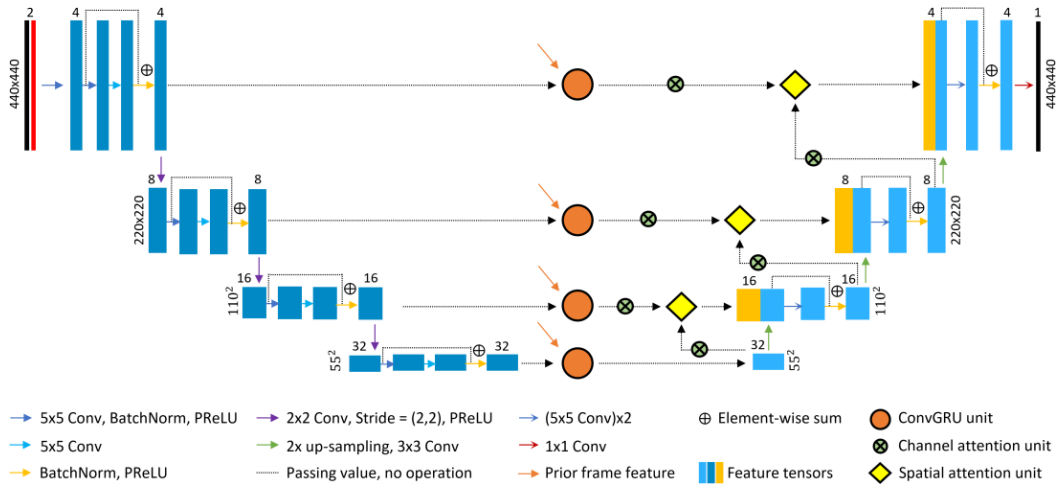

**Figure 3**. Spatiotemporal U-Net architecture. ConvGRU unit is taking encoded feature maps from the previous frame data. The spatial attention unit tries to utilize deeper level information with higher context value to guide segmentation of small targets, while the channel attention unit aims to adapt the weights for B-Mode and aggregated feature channel automatically.

## 2.3 Visibility-based surface ground truth label generation

For successfully training the proposed network on spine ultrasound images, we implemented a visibility-based ground truth label generator. First, after acquiring 3D volumetric scans of the vertebrae, we manually annotate the spinal surface ultrasound responses where high confidence is ensured. Then we utilize an iterative closest point (ICP)[11] algorithm to register the annotated surface points to the CT model of the spine. With the registered US-CT data, we can utilize the pose information of each contributing slice of the US volume and create ray beams corresponding to the imaging poses. Finally, we take the points when the rays first hit the CT spine surfaces and apply a 2D Gaussian filter to represent the uncertainty in registration. An output example can be seen in Fig. 4 (Ground Truth).

## 3. EXPERIMENTS AND RESULTS

### 3.1 Experiment environment setup

Two sets of CT data of patients with no spine abnormalities were obtained from public dataset[12]. Vertebrae L3-L5 were manually segmented and 3D printed. The spine models were embedded within the container filled with gelatin. Metamucil fiber is also added for creating realistic soft tissue scattering appearance, an example can be seen in Fig. 4 (BMode). In total, 3367 frames of images were acquired from 4 sets of scan experiments: each set of scans is located on different sides of different spine models, and each scan contains 8 linear sweep motion with distinct imaging angle by the patch scanner. For the following experiments, 70% of images were used for training and 30% for testing.

### 3.2 Result and Discussion

To evaluate the effectiveness of each proposed mechanism used in our surface estimator, we have conducted the ablation study: the controlled variable experiment results are summarized in the Tab. 1. Key observations include:

(1) *Exp1 vs. Exp2*: Our weighted Dice (w-Dice) loss (using label/prediction intensity as weights) outperforms the traditional weighted Cross-Entropy (w-CE) loss.
(2) *Exp1 vs. Exp3*: Compared to a pure CNN (original U-Net), networks with ConvGRU units can learn from both spatial and temporal context to make better estimation.
(3) *Exp1 vs. Exp4*: If we align the RNN reset length (# of frames to run before reset) with patch scanning motion, i.e., sweep along same direction, the network can learn the repetitive anatomy and scanning pattern better, comparing to using a fixed length reset during training.
(4) *Exp4 vs. Exp5*: If the training and testing images coming from the same anatomy, the results will generally be better than testing on an unseen anatomy, but the difference is still minor. We will be making more phantoms with varying anatomy and enhanced realism to ensure network generalizability.
(5) *Exp4 vs. Exp6*: With the addition of our aggregated feature channel (with <1% increase of overall network size), the network can obtain a major improvement comparing to using B-Mode image only.

**Table 1**. Controlled variable experiments summary. The experiment group pairs include (1,2), (1,3), (1,4), (4,5), (4,6).

| Experiment ID | Loss Type | Network Type | RNN Reset Type | Test Data | Input Channel | Avg. Dice score |
|---|---|---|---|---|---|---|
| 1 | w-Dice | RNN | Fixed-length | Unseen image | BMode+Feature | 0.422 |
| 2 | w-CE | RNN | Fixed-length | Unseen image | BMode+Feature | 0.416 |
| 3 | w-Dice | CNN | N/A | Unseen image | BMode+Feature | 0.287 |
| 4 | w-Dice | RNN | Align-with-scan | Unseen image | BMode+Feature | 0.493 |
| 5 | w-Dice | RNN | Align-with-scan | Unseen anatomy | BMode+Feature | 0.436 |
| 6 | w-Dice | RNN | Align-with-scan | Unseen image | BMode only | 0.331 |

A qualitative result example of our surface estimator is also shown in Fig. 4. We can visually confirm the 3D vertebrae surface model is well recovered and can potentially be used to guide clinician to perform lumbar intervention.

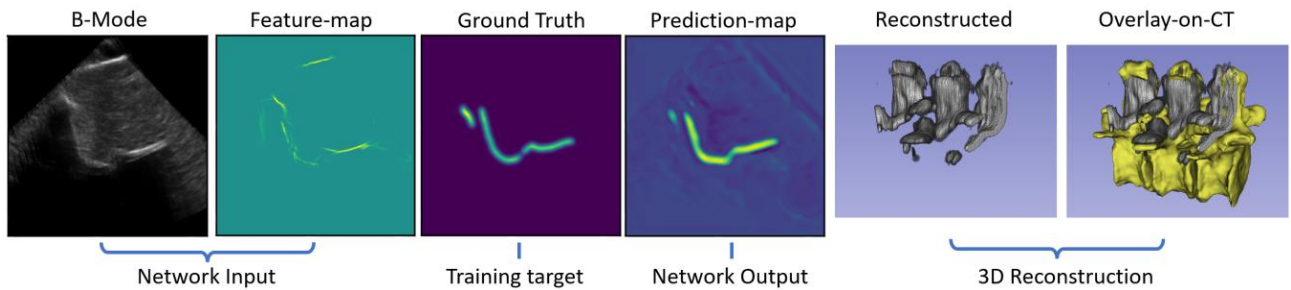

**Figure 4**. This sample is extracted from *Exp4*, the frame shown has achieved weighted Dice score of 0.58 for intuitive interpretation of our results in Tab 1. The reconstructed volume is visualized in 3D Slicer and overlaid on registered CT model of our spine phantom for comparison.

## 4. CONCLUSIONS

In this work we presented a method for estimating the complex shaped vertebra surface from ultrasound images acquired by a wearable patch-like ultrasound scanner. Our results showed that with spatiotemporal context learning, our surface estimator can achieve promising results for guided lumbar intervention using the reconstructed 3D model. In the future we will expand database for network training and integrate the wearable ultrasound imaging system with an augmented reality-based guidance system to evaluate overall system performance.